\documentstyle[epsf]{mn}

\title{M giants in MACHO, DENIS and ISOGAL}

\author[I.S. Glass, M.Schultheis]{I.S. Glass$^1$ and M. Schultheis$^2$ \\
$^1$South African Astronomical Observatory, PO Box 9, Observatory 7935, South
Africa (email: isg@saao.ac.za)\\$^2$Institut d'Astrophysique, 98 bis Blvd
Arago, Paris, France}

\date{Submitted 2002;
accepted 2002 August 1 }

\begin{document}

\maketitle

\begin{abstract}

A `complete' sample of 174 M giants classified by Blanco (1986) and
later than subtype M0 in the NGC\,6522 Baade's Window clear field has been
investigated to establish some general properties of cool Bulge stars.
Photometric information has been obtained from the MACHO database to search
for variablility and, where possible, to determine periods. Near- and mid-IR
magnitudes have been extracted from DENIS and ISOGAL.

Forty-six semi-regular (SR) variables and two irregular variables were found
amongst the 174. Many M5 and all stars M6 or later show variation,
whereas earlier subtypes (M1--M4) do not. The DENIS $I-J$ and $J-K_S$
colours and the luminosities of the M stars increase with M sub-class. $K$
tends to increase with log\,$P$ among the M-type SR variables.

Almost all the variables were detected at 7$\mu$m during the ISOGAL
programme. Excess radiation at 15$\mu$m, indicative of heavy mass-loss, is
associated with high luminosity and late spectral type. The limit of
sensitivity of the ISOGAL survey was such that the non-variables were not
detected.

Four probable M stars not listed by Blanco (1986), two of which are
semi-regular variables, were detected by ISOGAL.

In the case of doubly-periodic SR variables, the longer periods have
$K$-mags which place them close to the `D' line of Wood (2000) in a $K$, log
$P$ diagram. The unusual MACHO light curve of one particular star, Blanco
26, shows the commencement of a long-period variation with an anomalously
short and sharp event and appears to rule out a pulsational model for this
phenomenon.

\end{abstract}

\begin{keywords}
Stars:variables:others, stars: AGB

\end{keywords}

\section{Introduction}

The Baade's Window fields in the inner Bulge of the Milky Way galaxy offer
the opportunity of studying a sample of galactic stars at a well-defined
distance from the sun and with relatively low interstellar absorption. 

The mira variable content of the NGC\,6522 and Sgr I Baade's Windows has
been surveyed by Lloyd Evans (1976). A recent census of miras in Sgr\,I,
which includes the results of blue and near-infrared photographic surveys as
well as long-period stars found during the IRAS mid-infrared survey, has
been given by Glass et al (1995). In the present work, the general
population of M giants, including variables of lower amplitude than the
miras, is studied in the NGC\,6522 Window.  A `complete' sample of M stars
from M1 to M8 in an annular field of 42 arcmin$^2$ area surrounding the
NGC\,6522 cluster was obtained by Blanco (1986) using a grism attached to
the prime focus camera of the CTIO 4m telescope. It should be noted that the
classification into M sub-types by Blanco (1986) is not quite on the
Morgan-Keenan (MK) system, in that stars of M2 to M6 were stated to have
been classified 1-2 subtypes too late by Terndrup, Frogel and Whitford
(1990; see also below). Spectroscopy of all Blanco's stars of subtype M6 or
later was obtained by Sharples, Walker \& Cropper (1990), who tabulated
their radial velocities and their TiO (8415) and CaII (8662) line strengths.

Unlike the solar neighbourhood and the Magellanic Clouds, the galactic
centre fields do not contain late-type C-rich giants. The relatively small
Blanco (1986) field also does not include any miras. Lloyd Evans' (1976)
figures for the surface density of miras suggest that only about one could
have been expected.

The ISOGAL survey (Omont et al., 2002) at 7 and 15$\mu$m of sample fields
along the Galactic Plane and within the Galactic Bulge included a `fiducial'
field around the globular cluster NGC\,6522. Glass et al.\ (1999) showed that
all stars later than M6 were detected, whereas M1 types were not
seen at all. It was found that many AGB stars besides the mira variables
possess excesses at 15$\mu$m, indicating that they are losing mass.

Glass, Alves et al.\ (2000; see also Alard et al., 2001) showed that nearly
all the stars that were detected by ISOGAL in the NGC\,6522 and Sgr\,I
Baade's windows could be identified with late-type variables in the MACHO
database. They studied approximately 300 stars that were seen in both the
MACHO $v$ and $r$ colours as well as in both the ISOGAL bands at 7 and
15$\mu$m. Almost all the non-miras were found to be semi-regular variables,
with mass-loss rates that ranged from undetectable to $ \sim5 \times$
10$^{-7}$
$M_{\odot}$yr$^{-1}$. Semiregular variables were seen to outnumber miras by
at least 20:1 in these fields. They had periods ranging from 10-200 days and
amplitudes which reached $\sim$0.3 mag at 100d period. A period of 70 days
or more was found to be a necessary, though not a sufficient, condition for
detectable mass-loss.

Schultheis and Glass (2001) examined the same sample of stars in the near
infrared period-magnitude, period-colour and colour-magnitude diagrams
derivable from the DENIS $IJK_S$ dedicated survey of the galactic bulge
(Simon et al., 2002, in preparation) and the 2MASS $JHK_S$ survey.  The
Semi-regular variables inhabit the upper end of the $J-K_S, K_S$
colour-magnitude diagram and their colours and magnitudes are seen to
increase in a general way with period. The well-known separation between
miras and late-type giants in the $J-H, H-K$ diagram was shown to be
primarily caused by the effect of water vapour absorption on the
$J-H$ colour index.

In this paper, we examine the complete Blanco (1986) sample of stars in
the NGC\,6522 field classified M1 or later for variability in the MACHO
database to find the dependence of variability on spectral type.
Because many of these stars fall within the area surveyed by ISOGAL they
could also be examined for mass-loss. The previous work (Alard et al, 2001;
Schultheis and Glass, 2001) dealt exclusively with stars detected in both MACHO
($r$ and $i$) and both ISOGAL (7$\mu$m and 15$\mu$m) bands and was heavily
weighted towards stars with mass-loss. There are only 23 stars in common.

\section{Extraction of MACHO data}

\subsection{Positions}

Blanco (1986) lists 174 stars of types M1--M8 near NGC\,6522. Using his
published finding chart, a Digitized Sky Survey FITS file of the area,
containing World Coordinates, was used to extract positions for each object.
The MACHO on-line database was then searched, over a small radius of a few
arcsec surrounding each position, for red stars. The positions of the
candidates, as given in the MACHO database, are included in Table 1. 
An overlay was created to check the positions against the Blanco chart for
errors.

\begin{figure*} 
\begin{minipage}{17.5cm} 
\epsfxsize=17.5cm 
\epsffile[28 152 539 786]{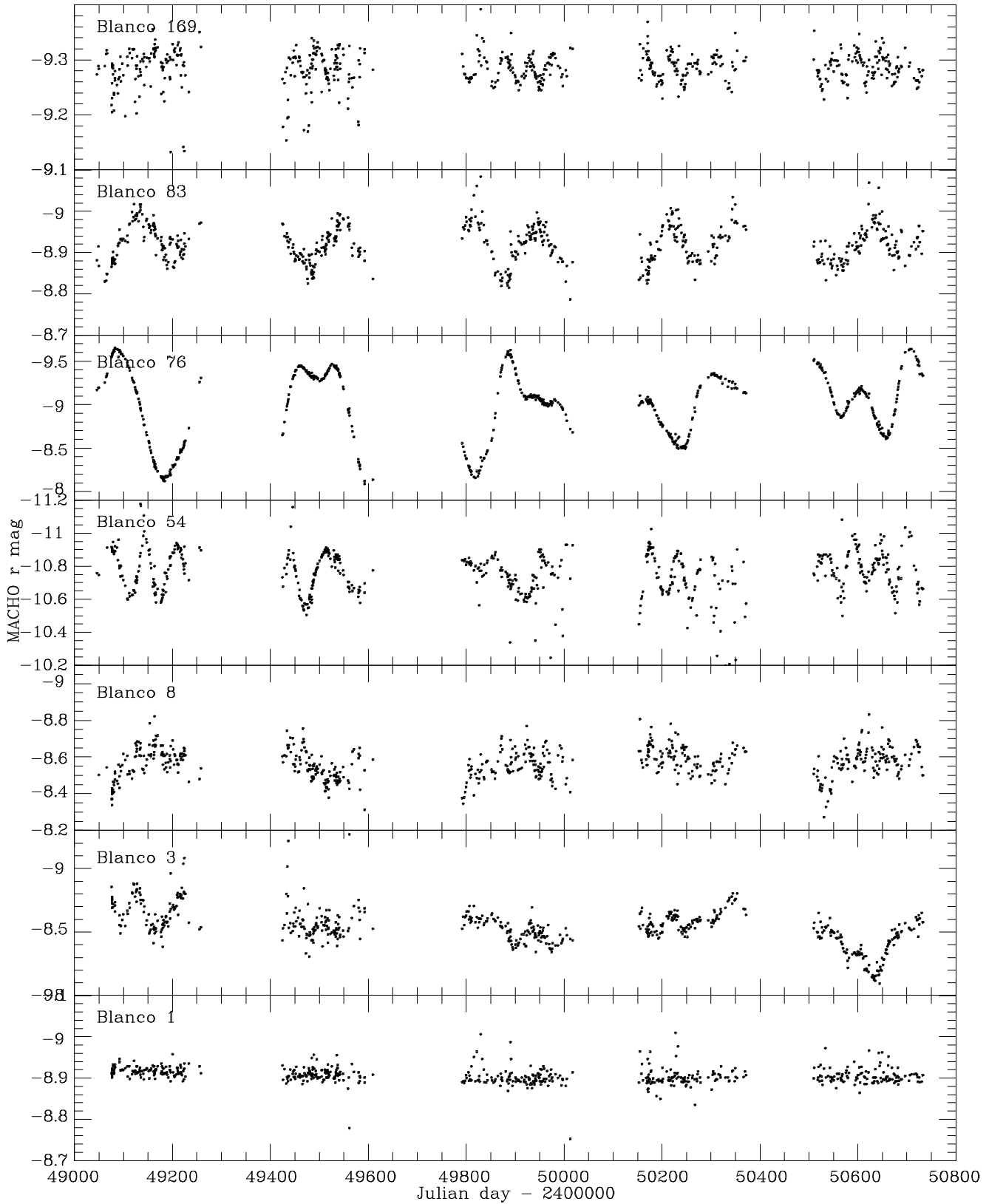} 

\caption{Various MACHO (instrumental) $r$-band light curves. (1) Typical
non-variable, (3) aperiodic variable, (8) typical variable with long [246d]
and short [31d] periods, (54) classified by Blanco as M1 but possibly a
later-type or composite [60d], (76) largest-amplitude variable [207d], (83)
Regular variable of long period [135d], (169) Regular `SR' variable of
shorter period [48d]. The magnitudes are on the natural system of MACHO,
i.e., they have not been corrected for zero points.}

\end{minipage}
\end{figure*}

\begin{table}
\caption{Baade's NGC6522 window M stars}
\begin{tabular}{llllll}
Bl. &  MACHO$^1$    &RA$^2$ & Dec & M & Per(s) Ampl(s) \\
No. &  119.20   &     &             & ty.  & \\

1  &  091.3861 &17.5& -30 01 26.01& 1  &                        \\
2  &  091.3867 &17.7& -30 02 31.47& 4  &                        \\
3  &  091.3890 &17.8& -30 02 29.50& 6.5&  irr  0.7  n.i.D.      \\
4  &  091.3853 &18.1& -30 03 11.02& 6  &  11.4 0.05 :           \\
5  &  091.3882 &18.2& -30 01 28.58& 1  &                        \\
6  &  091.3863 &18.3& -30 01 50.84& 1  &                        \\
7  &  091.3921 &18.4& -30 02 35.20& 3  &                        \\
8  &  091.3885 &18.7& -30 02 19.89& 6  &  246 0.2 30.8 .05      \\
9  &  091.3849 &19.3& -30 03 18.91& 3  &                        \\
10 &  091.3864 &19.5& -30 03 15.79& 1  &                        \\
11 &  091.3904 &20.0& -30 01 10.36& 2  &                        \\
12 &  091.3841 &20.2& -30 00 41.23& 3  &  39.1 0.15             \\
13 &  091.3857 &20.6& -30 01 46.34& 1  &                        \\
14 &  091.3891 &21.0& -30 02 43.20& 4  &                        \\
15 &  091.3964 &21.2& -30 01 50.45& 1  &                        \\
16 &  090.3749 &21.5& -30 04 19.18& 2  &                        \\
17 &  091.3854 &21.5& -30 03 20.66& 4  &                        \\
18 &  091.3881 &21.8& -29 59 46.99& 1  &                        \\
19 &  090.3744 &21.9& -30 03 52.34& 3  &                        \\
20 &  092.4036 &21.9& -29 59 23.42& 3  &                        \\
21 &  091.3856 &21.8& -30 01 21.64& 3  &                        \\
22 &  090.3720 &22.0& -30 04 29.95& 4  &                        \\
23 &  091.3858 &22.3& -30 00 39.05& 4  &                        \\
24 &  091.3839 &22.3& -30 02 56.05& 6  &  51.4 0.2              \\
25 &  091.3844 &22.8& -30 03 07.83& 2  &                        \\  
26 &  091.3879 &23.0& -30 03 19.59& 6  &  344 0.35, 27 0.1      \\
27 &  221.173  &23.7& -30 02 10.50& 2  &                        \\
28 &  221.80   &23.9& -30 00 05.84& 5  &  48.6 0.1, 399.1 0.6   \\
29 &  221.75   &23.9& -30 03 20.08& 2  &                        \\
30 &  221.111  &23.9& -30 00 47.98& 2  &                        \\
31 &  222.2570 &23.9& -29 59 25.74& 6.5&  66.4 0.3              \\
32 &  221.216  &24.1& -30 02 05.12& 1  &                        \\
33 &  221.118  &24.3& -30 01 42.20& 2  &                        \\
34 &  220.61   &24.4& -30 04 15.87& 6.5&  34.3 0.15             \\
35 &  220.102  &24.5& -30 04 39.10& 6  &  15.2 0.1              \\
36 &  222.2522 &25.1& -29 59 16.8 & 6  &  66.2 0.1              \\
37 &  221.109  &25.1& -30 01 21.03& 1  &                        \\
38 &  221.104  &25.2& -29 59 48.3 & 6  &  25.7 0.2              \\
39 &  222.2546 &25.7& -29 58 47.0 & 6  &  very noisy            \\
40 &  221.165  &26.9& -30 01 45.48& 2  &                        \\
41 &  220.151  &26.9& -30 04 25.82& 1  &                        \\
42 &  221.84   &27.1& -30 03 08.49& 2  &                        \\
43 &  221.190  &27.1& -30 02 10.63& 1  &                        \\
44 &  222.2505 &27.1& -29 59 07.63& 1  &                        \\
45 &  221.126  &27.2& -30 01 03.27& 6  &  749 0.3 71.2 0.2      \\
46 &  220.86   &27.3& -30 04 34.4 & 4  &                        \\
47 &  221.55   &27.4& -30 02 25.76& 5  &  16.3 0.1              \\
48 &  220.98   &27.6& -30 04 21.92& 1  &                        \\
49 &  221.93   &27.7& -30 00 56.34& 1  &                        \\
50 &  221.112  &27.7& -30 00 17.82& 1  &                        \\
51 &  222.2534 &28.2& -29 58 58.43& 4  &                        \\
52 &  220.129  &28.2& -30 04 10.60& 1  &                        \\
53 &  221.155  &28.8& -30 02 27.51& 7  &  20.3 0.15             \\
54 &  221.45   &29.2& -30 02 48.98& 1  &  60.3 0.4 composite?   \\
55 &  221.263  &29.2& -30 03 23.75& 1  &                        \\
56 &  220.74   &29.3& -30 05 43.69& 1  &                        \\
57 &  222.2540 &29.3& -29 59 39.73& 7  &  aperiodic ~0.4 amp    \\
58 &  220.63   &29.5& -30 05 34.04& 5  &                        \\
59 &  220.176  &29.6& -30 05 08.69& 1  &                        \\
60 &  221.64   &30.0& -30 03 14.90& 2  &                        \\
61 &  222.2501 &30.0& -29 58 21.47& 5  &  40.7 0.2              \\
62 &  222.5697 &30.5& -29 58 35.71& 2  &  19.6 .05 n.i.D.       \\  
63 &  221.81   &30.6& -30 00 50.97& 6  &  395 0.4, 73.9 0.2     \\
64 &  221.91   &30.7& -30 00 28.10& 3  &                        \\
65 &  222.2573 &31.0& -29 59 02.6 & 7  &  61.8 0.2              \\
\end{tabular}                                                             
\end{table}                                                               

\setcounter{table}{1}
\begin{table}
\caption{Baade's NGC6522 window M stars {\it contd.}}
\begin{tabular}{llllll}
Bl. &  MACHO$^1$    &RA$^2$ & Dec & M & Per(s) Ampl(s) \\
No. &  119.20   &     &             & ty.  & \\

66 &  221.44   &31.7& -30 00 44& 3  &                        \\
67 &  222.2541 &31.6& -29 59 11& 1  &                        \\
68 &  220.108  &31.9& -30 04 09& 2  &                        \\
69 &  221.56   &32.0& -30 00 29& 6  &  331   0.2 43.6 0.1    \\
70 &  220.81   &32.3& -30 04 44& 6  &  25.8  0.1             \\
71 &  221.87   &32.3& -30 03 32& 4  &  17.3  0.05            \\
72 &  222.2523 &32.8& -29 59 39& 3  &                        \\
73 &  220.192  &33.2& -30 05 08& 1  &                        \\
74 &  220.132  &33.2& -30 04 41& 3  &                        \\
75 &  220.89   &33.3& -30 05 23& 7  &  575, 0.1 70.0  0.15   \\
76 &  222.2518 &33.2& -29 59 11& 6.5&  207  1.0              \\
77 &  221.77   &33.7& -30 03 31& 3  &  50.7  0.02 :          \\
78 &  220.33   &33.7& -30 03 57& 2  &                        \\
79 &  222.2547 &33.9& -29 59 22& 1  &                        \\
80 &  220.65   &34.1& -30 05 17& 6  &  16.6  0.05            \\
81 &  221.178  &34.1& -29 59 59& 8  &  120.5 0.3             \\
82 &  221.298  &34.4& -30 00 17& 2  &                        \\
83 &  221.154  &35.0& -29 59 49& 5  &  135.5 0.1 regular     \\
84 &  220.31   &35.1& -30 05 25& 5  &                        \\
85 &  222.2536 &35.1& -29 59 09& 1  &                        \\
86 &  222.2499 &35.2& -29 58 13& 1  &                        \\
87 &  221.135  &35.3& -30 00 01& 1  &                        \\
88 &  220.119  &35.7& -30 04 39& 1  &                        \\
89 &  220.126  &35.6& -30 04 00& 1  &                        \\
90 &  220.42   &35.8& -30 05 23& 2  &                        \\
91 &  222.2548 &35.8& -29 58 35& 1  &                        \\
92 &  222.2557 &36.0& -29 58 58& 6.5&  28.5 0.1              \\
93 &  221.97   &36.0& -30 00 22& 1  &                        \\
94 &  221.138  &36.5& -30 00 40& 5  &                        \\
95 &  221.85   &36.9& -30 03 31& 1  &                        \\
96 &  222.2503 &37.0& -29 58 47& 4  &                        \\
97 &  220.79   &38.4& -30 04 16& 1  &                        \\
98 &  220.122  &38.9& -30 04 38& 2  &                        \\
99 &  221.228  &38.9& -30 03 35& 2  &                        \\
100&  221.103  &39.0& -29 59 53& 1  &                        \\
101&  222.2502 &39.0& -29 58 26& 4  &                        \\
102&  221.121  &39.3& -29 59 50& 1  &                        \\
103&  222.2526 &39.6& -29 58 24& 1  &                        \\
104&  221.114  &39.7& -30 00 52& 2  &                        \\
105&  222.2511 &39.7& -29 58 48& 3  &                        \\
106&  221.136  &40.5& -30 03 28& 2  &                        \\
107&  222.2532 &40.4& -29 59 11& 1  &                        \\
108&  221.158  &40.8& -29 59 58& 1  &                        \\
109&  220.58   &41.2& -30 05 03& 1  &  noisy .3              \\
110&  221.62   &41.3& -30 03 00& 3  &                        \\
111&  351.88   &41.3& -29 59 53& 1  &                        \\
112&  352.2211 &41.5& -29 59 21& 4  &                        \\
113&  350.108  &41.8& -30 04 39& 1  &  noisy .2              \\
114&  351.212  &42.2& -30 02 39& 2  &                        \\
115&  351.98   &42.2& -30 01 13& 2  &                        \\
116&  350.70   &42.5& -30 04 59& 4  &                        \\
117&  351.136  &42.6& -30 03 37& 2  &                        \\
118&  351.22   &42.6& -30 02 50& 1  &                        \\
119&  351.36   &42.6& -30 00 07& 5  &  18.5 0.05             \\
120&  352.2218 &42.7& -29 58 59& 3  &                        \\
121&  352.2247 &43.2& -28 58 42& 1  & n.i.D.                 \\
122&  351.76   &43.4& -30 03 24& 1  &                        \\
123&  351.150  &43.5& -29 59 57& 1  &                        \\
124&  350.124  &43.8& -30 05 17& 7  &  561 0.2, 71.6 0.1     \\
125&  351.135  &43.8& -30 00 45& 2  &                        \\
126&  351.20   &44.0& -30 02 21& 6  &  127.2 0.5,            \\
   &           &    &          &    &  80.0 0.15             \\
127&  350.250  &44.3& -30 03 43& 7  &                        \\
128&  352.2220 &44.5& -29 59 29& 2  &                        \\
129&  350.34   &45.1& -30 04 57& 3  &                        \\
\end{tabular}                                                             
\end{table}                                                               

\setcounter{table}{1}
\begin{table}
\caption{Baade's NGC6522 window M stars {\it contd.}}
\begin{tabular}{llllll}
Bl. &  MACHO$^1$    &RA$^2$ & Dec & M & Per(s) Ampl(s) \\
No. &  119.20   &     &             & ty.  & \\

130&  352.2206 &45.1& -29 58 47& 3  &                        \\
131&  350.42   &45.5& -30 05 18& 3  &                        \\
132&  351.46   &45.3& -30 01 33& 5  &  35.6 0.15             \\
133&  351.14   &45.5& -30 03 00& 4  &  bright, noisy         \\
134&  350.75   &45.5& -30 04 33& 6  &  529 0.1, 53.4 0.1     \\
135&  351.217  &45.2& -30 00 40& 1  &                        \\
136&  351.153  &45.5& -30 00 09& 3  &                        \\
137&  351.35   &46.2& -30 02 24& 6.5&  188.1 0.3             \\
138&  352.2239 &46.0& -29 59 13& 7  &  173.5 0.3             \\
139&  350.57   &46.5& -30 03 54& 5  &  irr var?, $<$~0.05    \\
140&  351.66   &46.4& -30 00 48& 2  &                        \\
141&  351.100  &46.4& -30 01 38& 2  &                        \\
142&  351.64   &46.6& -30 02 27& 6.5&  23.7 0.15             \\
143&  351.103  &46.8& -30 02 45& 4  &                        \\
144&  351.48   &47.0& -30 02 02& 3  &                        \\
145&  351.111  &47.4& -30 02 10& 1  &                        \\
146&  351.79   &47.4& -30 03 37& 6.5&  24.1 0.1              \\
147&  351.84   &47.4& -30 01 22& 2  &                        \\
148&  350.109  &47.6& -30 04 06& 5  &                        \\
149&  352.2221 &47.6& -29 59 16& 4  &                        \\
150&  351.140  &48.9& -30 01 33& 1  &                        \\
151&  351.42   &48.0& -30 00 38& 5  &                        \\
152&  351.39   &48.2& -30 01 03& 3  &                        \\
153&  351.83   &48.6& -30 00 33& 2  &                        \\
154&  351.63   &48.5& -29 59 47& 6.5&  219.3 0.2,            \\
   &           &    &          &    &  67.9 0.15             \\
155&  352.2197 &49.1& -29 59 35& 3  &                        \\
156&  351.112  &49.6& -30 03 41& 1  &                        \\
157&  352.2241 &49.4& -29 59 41& 2  &                        \\
158&  351.62   &49.8& -30 01 35& 3  &                        \\
159&  351.41   &49.9& -30 02 23& 1  &                        \\
160&  351.15   &49.9& -30 03 00& 2  &                        \\
161&  351.28   &50.1& -30 03 14& 5  &  22.7 32.3             \\
   &           &    &          &    &  eq. amps., $\sim$ 0.1 \\
162&  351.65   &50.5& -30 02 20& 4  &                        \\
163&  351.21   &50.9& -30 01 52& 3  &  23.2 0.05             \\
164&  351.40   &51.1& -30 00 14& 2  &                        \\
165&  351.131  &51.1& -30 00 37& 1  &                        \\
166&  351.34   &51.2& -30 01 43& 2  &                        \\
167&  351.55   &51.2& -30 01 21& 2  &                        \\
168&  351.134  &51.5& -30 01 00& 1  &                        \\
169&  351.57   &52.0& -30 02 02& 1  &  48.0 0.05 regular     \\
170&  351.138  &52.1 & -30 03 03& 2  &                        \\
171&  351.29   &52.2& -30 03 33& 5  &  471 0.1, 34.4 .06,    \\
   &           &      &             &    &  25.6 .05              \\
172&  351.37   &52.6& -30 01 23& 6  &  44.7 0.05             \\
173&  351.121  &52.9& -30 01 59& 6  &  31.9 0.15             \\
174&  351.80   &53.4& -30 02 08& 2  &                        \\
\end{tabular}          

Notes:

\noindent Positions are epoch 2000

\noindent $^1$ The full MACHO number is prefixed by 119.20, e.g.
119.20091.3861 for Blanco 1.

\noindent $^2$ Seconds of Right Ascension. Add 18$^{\rm h}$ 03$^{\rm m}$
                                                 
\noindent n.i.D. = Not in DENIS
  
\end{table}                                                               
       
\subsection{Variability}

The $r$ light curve and the Fourier spectrum of each object were plotted.
Each light curve was examined by eye as well as by computer and generally
any variability detected by machine could also be seen.  The apparent
`noise' on the light curves was not constant, probably due to crowding, but
it is believed that periodic or near-periodic variations of 0.03 mag or more
peak-to-peak would all have been detected. Because of the seasonal nature of
the data, the analysis may have been somewhat less reliable at longer periods
($>$ 100d).

The Alard et al (2001) data cover six observing seasons, while the initial
web-based set of data used here covers only five. The periods given by Alard
et al (2001) represent the highest amplitudes seen in the summed Fourier
spectra for each individual season (for log $P$ $<$ 2.2), whereas in this
work the entire series of observations for each star was analysed at the
same time. The photometry was plotted according to phase for the main period
found and the average amplitude at that period was estimated by eye. It
should be accurate to about 0.1 mag.

Generally, more attention has been paid to secondary periods and long-period
components than in the Alard et al work. In some cases, the data were
pre-whitened by removing the main component of variation in order to make
the minor components more easily recognizable. This procedure is not very
satisfactory when the amplitudes are not constant. The resultant periods and
amplitudes are given in Table 1. 

Altogether 48 Blanco stars were found to be variable (see Table 2).  The
early (Blanco M1--M4) subtypes are only rarely seen to vary at the level we
can detect whereas variability becomes ubiquitous for the later subtypes
(M5--M8). Of the variables, 46 showed some degree of coherence throughtout
the series of observations and can therefore be considered to be SRa type,
as judged by eye from light curves phased according to the derived periods.
Only two seemed to be completely lacking in long-term coherence for any
period. The reality of the shortest period given in Table 1, 11.4 days for
Blanco 4, is somewhat doubtful, but other short periods around 15 days seem
to be well-established.

Various previous studies have sought to determine the dependence of
variability on spectral type. Jorissen et al (1997) suggested that
variability at a low level is ubiquitous in late-K and early-M giants. Henry
et al (2000) demonstrated short-term variability at about the 0.01--0.02 mag
level in M0 giants. This level is probably not detectable in the MACHO data.
Koen \& Laney (2000) examined the Hipparcos data for nearby late-type stars
in the solar neighbourhood and found evidence for variability, usually with
amplitude $<$ 0.1 mag, in many early-M type stars. However, Kerschbaum,
Lebzelter \& Lazaro (2001) have observed some of these from the ground and
question whether the shortest periods seen (i.e. $<$ 10d) might not be
artefacts of the Hipparcos data.
  
\subsubsection{Notes on individual stars}

Most of the periods agree quite well with the Alard et al (2001) results. The
differences that occur appear mainly to be due to the additional season
of data in the latter and appear to reflect the irregularity of the
variations that are being measured. The following are the most striking
examples:

\vspace{1mm}

\noindent Blanco 3. This is a highly irregular star (see Fig 1) in which
Alard et al found a period of 115d to have the highest amplitude. It does
not appear to show any persistent period.

\noindent Blanco 4. Alard et al found a period of 80 days with an amplitude
0.1 mag. The present data may support a period of 11.39d with a very small
amplitude of about 0.02 mag.

\noindent Blanco 38. The Alard et al period of 37.4 d has a slightly smaller
amplitude than the period of 25.7d, found in this work.

\noindent Blanco 39. The MACHO website programmes did not appear to extract
the light curve of this star correctly; the data appearing to be very
noisy. The Alard et al value for the period is used.

\noindent Blanco 57. This very irregular variable shows periods of 502, 171
and 97 days, none of which agree with the Alard et al period of 118.9d. 

\noindent Blanco 138. Alard et al give 119d.

\vspace{1mm}

\subsection{$r$ amplitudes}

The average $r$ amplitudes of the variables increase with log $P$ (fig.\ 2).
In this diagram, when stars have both short and long periods, the short
periods follow the trend of the singly-periodic variables.

\begin{figure}                                                            
\epsfxsize=8.2cm                                                          
\epsffile[25 282 539 786]{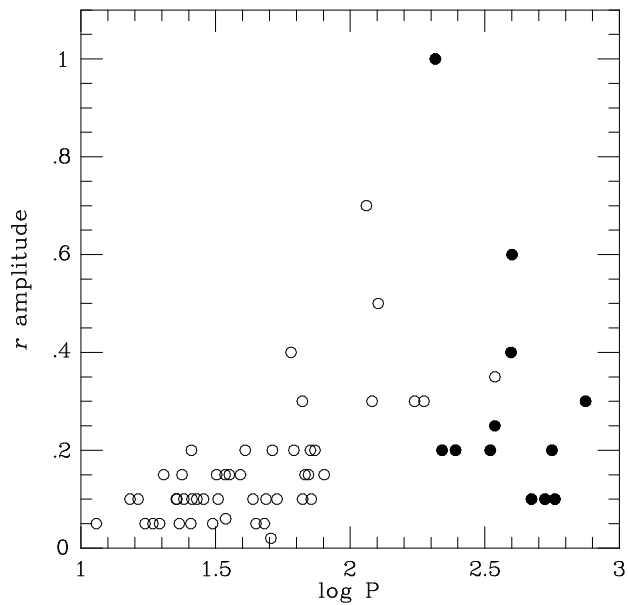}                                           
\caption{MACHO $r$ amplitudes as a function of log $P$. Some stars have both
short and long periods and the longer periods are represented here by filled
circles.}
\end{figure}                                                              
                           
\section{Cross-correlation with DENIS}

The DENIS database of $IJK_S$ magnitudes was searched for the 174 Blanco
stars. A search radius of 3$''$ was used (the astrometric precision of the
DENIS positions is expected to be $\sim$ 1$''$). Counterparts were not
found for Blanco 2 (very close to Blanco 3 in position) and Blanco 121. The
latter was classified M1 by Blanco (1986).

Figure 3 shows a plot of DENIS $K_S$ mag vs M subtype for the stars that
were identified. The variables are also indicated. There is a steady
increase in average $K_S$ flux, and therefore luminosity, with spectral
sub-type in this sample, from the early to the later types. A few outlying
stars may be in front of or behind the main Bulge population. The three
brightest stars of type M6 and the brightest of type M6.5 have $I$ $\leq$
11.8, according to Sharples et al (1990), which places them within a group
having low velocity dispersion that lies in the foreground.

Stars of Blanco (1986) type M6 or later were all found to be variable, as
were the most luminous of the M5 group. The earlier types were, with only a
few exceptions, not seen to vary. Amongst the exceptions, Blanco 54
(classified M1), appears more luminous than others of the same spectral type
in $K_S$ and also in Blanco's (1986) original list of $V$ mags.  Its light
curve (fig.\ 1) is suggestive of a later type than M1. Its DENIS
$(I-J,J-K_S)_0$ colours of (1.23, 1.01) suggest M2--M3. It may a
superimposition of an early and a late-type star. 

Variability can be seen to be associated with both high luminosity and late
spectral type. It is not clear which of these should be regarded as the most
fundamental. The tip of the Red Giant Branch is usually considered to be $K$
$\sim$ 8.2 (e.g., Tiede, Frogel \& Terndrup, 1995; Omont et al, 1999) and it
is clear that variability extends to fainter magnitudes than this. However,
AGB stars also occur at these luminosities and cannot be distinguished from
RGB stars.

\begin{table}
\caption{Variability and Blanco spectral sub-class}
\begin{tabular}{lll}
Blanco class & not variable & variable\\
M1   & 51  & 2$^1$ \\ 
M2   & 34  & 1 \\
M3   & 20  & 3 \\
M4   & 15  & 1 \\
M5   & 5  & 9 \\
M6   & 0 & 16$^2$ \\
M6.5 & 0  & 9 \\
M7   & 1  & 6 \\
M8   & 0  & 1 \\

\end{tabular}

\noindent

$^1$Includes Blanco 54 (possible a composite object of M1 + something else).

$^2$Includes Blanco 39 (noisy), which may or may not be variable.

\end{table}

\begin{figure} 
\epsfxsize=8.2cm 
\epsffile[25 282 539 786]{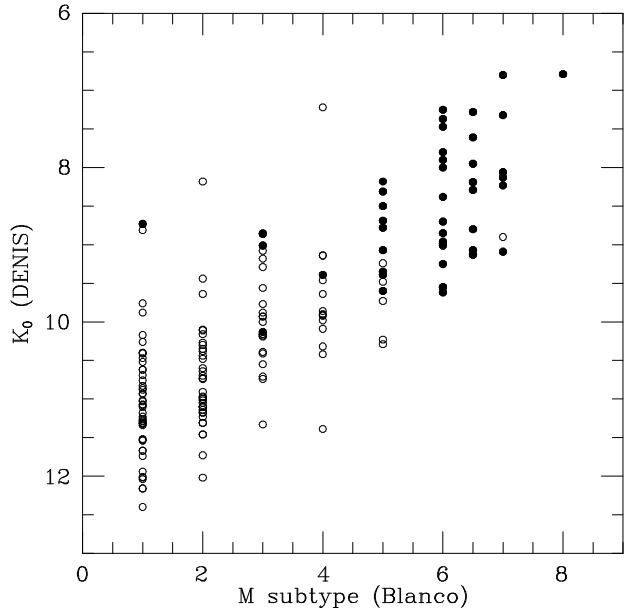}
\caption{DENIS $K_S$ mag vs spectral type for the complete sample. Known
variables are denoted by filled circles. No corrections for interstellar
absorption have been applied}
\end{figure}                                                              

The relation of DENIS $I-J$ and $J-K_S$ colours to Blanco (1986) spectral
type, at least for the M giants in Baade's NGC\,6522 Window, is shown in Fig
4. The colours have been corrected for $A_V$ = 1.45 mag (see Glass et al,
1999), or $E_{I-J}$ = 0.50 and $E_{J-K_S}$ = 0.23. A strong increase is
noticeable in $I-J$ colour towards later spectral types. The average
de-reddened colours are tabulated in Table 3.

\begin{figure}
\epsfxsize=8.2cm
\epsffile[28 0 539 786]{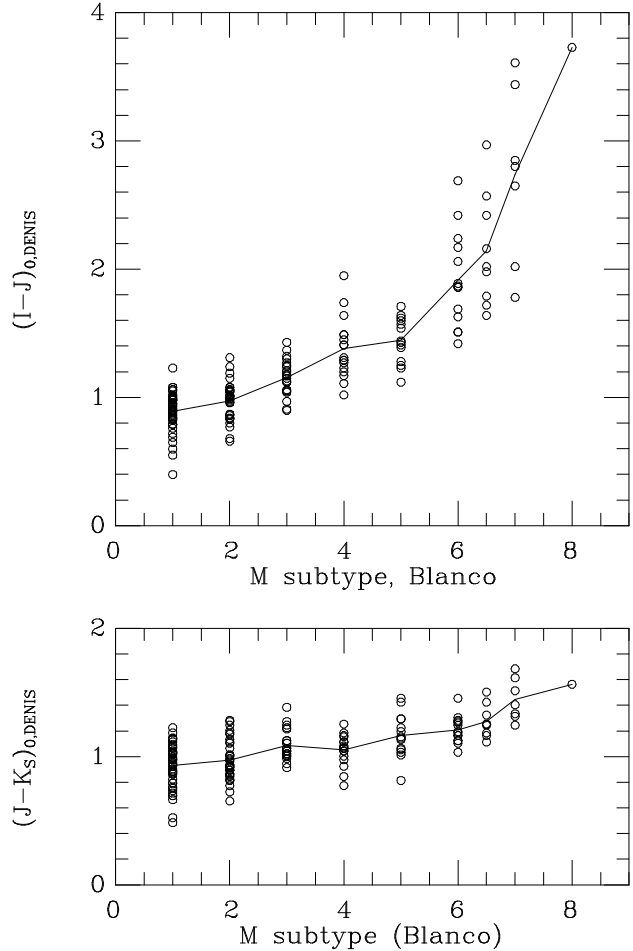}
\caption{DENIS colours vs Blanco (1986) spectral types for Baade's Window
stars. The lines represent the average values. See text concerning relation
of Blanco to MK spectral types. Interstellar absorption corrections
corresponding to $A_V$ = 1.45 mag have been applied.}
\end{figure}

A comparison can be made between the average DENIS colours for each Blanco
spectral type and synthetic colours for stars with MK spectral types that
have been measured spectrophotometrically by Lan\c{c}on and Wood (2000). Fig
5 shows the result. The two sets of stars for the most part occupy the same
locus, but the Lan\c{c}on and Wood colours are somewhat redder than the
Blanco ones for earlier spectral sub-types. As mentioned, Terndrup et al
(1990) stated that subtypes from M2 to M6 were classified 1-2 subtypes too
late, but we now see that this problem extends from M1 to M6. Stars with the
latest spectral subtypes of M7 and M8, although agreeing in
$I-J$, do not have the same $J-K_S$; this effect may be due to increased
mass-loss from the Bulge stars when compared to the Solar neighbourhood.

\begin{figure}
\epsfxsize=8.2cm
\epsffile[28 410 539 786]{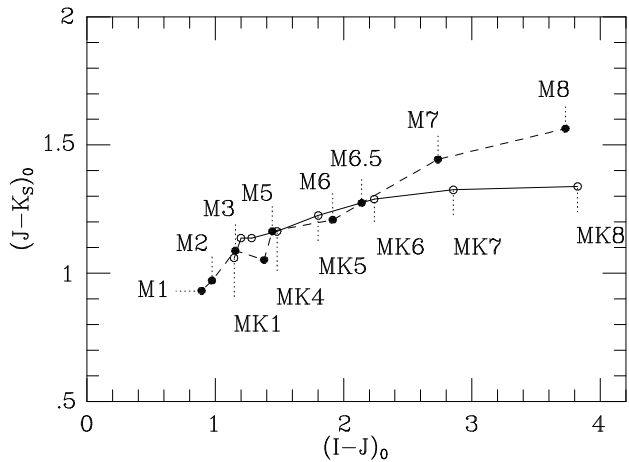}
\caption{Comparison of average DENIS colours of stars having Blanco (1986)
spectral types (M1--M8, dashed line) with synthetic DENIS colours
derived from spectrophotometric measurements of solar neighbourhood stars
with MK types (Lan\c{c}on and Wood, 2000; MK1--MK8, continuous line).}
\end{figure}

\subsection{Period-$K_S$ trend for NGC\,6522 SRVs}

The $K_S$ mag, log $P$ diagram for the variables is shown in Fig 6. Several
stars have two periods (or even more). In these cases, the longer periods are
shown as filled circles.

There is an obvious tendency of the $K_S$ mag to increase with period, if
the single-period stars and the short periods of the double-period stars are
considered.

\begin{table}
\caption{Average de-reddened DENIS colours}
\begin{tabular}{llll}
Blanco M subclass & no. & $(I-J)_0$ & $(J-K_S)_0$\\
  1    &47  &  0.893 & 0.932 \\
  2    &35  &  0.974 & 0.971 \\
  3    &23  &  1.156 & 1.088 \\
  4    &17  &  1.380 & 1.052 \\
  5    &14  &  1.444 & 1.164 \\
  6    &14  &  1.915 & 1.208 \\
  6.5  & 9  &  2.140 & 1.274 \\
  7    & 7  &  2.735 & 1.444 \\
  8    & 1  &  3.729 & 1.564 \\
\end{tabular}
\end{table}

\begin{figure}                                                            
\epsfxsize=8.2cm                                                          
\epsffile[25 282 539 786]{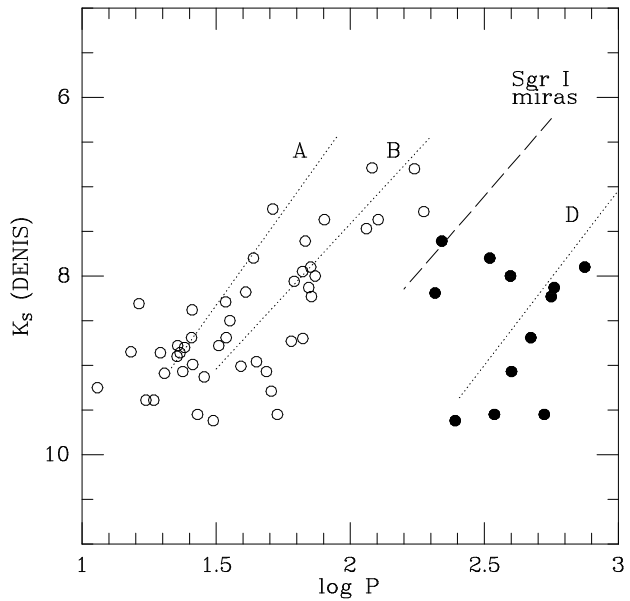}                                           

\caption{DENIS $K_S$ mag vs log P for the variable stars in the sample. Some
stars have both a long and a short period and the longer periods are
represented here by filled circles.  The dashed line is the
relation for Sgr I miras from Glass et al (1995). Wood's (2000) lines A, B
and D for SRVs in the LMC are also shown, adjusted for differences in
reddening ($\Delta K_S$ = 0.13) and distance modulus ($\Delta$dm = 3.9).
Wood's line C is similar to that for the Sgr I miras.}

\end{figure}                                                              

Wood (2000) has found that the SRVs in the LMC fall on a number of parallel
tracks in the $K_S$, log $P$ diagram (see fig.\ 6). Similar results have been
reported by Cioni et al (2001) using DENIS and EROS results. 
Wood (2000) interprets his tracks in terms of pulsation modes, where the
Mira sequence is the fundamental and the two sequences at shorter periods
are overtones. Track D (see next section) is thought to be an effect arising
from a close companion.

Because of the depth of the Bulge in the line of sight, any intrinsic
tendency of its SRVs to lie along Wood's sequences may not be obvious. For
example, the miras in the nearby Sgr I Baade's Window of the Bulge show a
dispersion from the $K_S$, log $P$ relation of 0.35 mag, which is much
greater than the 0.13 mag found for O-type miras in the Large Magellanic
Cloud (Glass et al, 1995, Glass in preparation). The increased scatter has
been ascribed, at least in part, to the finite thickness of the Bulge.

\subsection{The long periods}

The longer periods of the doubly-periodic SRVs are close to Wood (2000)'s
line D, which is followed by the longer periods of the similarly behaved AGB
stars in the LMC. The nature of the variation giving rise to this line
is not yet understood. It cannot be a pulsational mode if the mira sequence
is that of fundamental mode pulsators. Binarity may be involved as a
mechanism because the periods appear to be stable and about appropriate to
the orbit of a cool companion or giant planet in the outer part of the
atmosphere of a late-type M giant. Some evidence for this possibility, based
on radial velocity measurements, has been presented by Wood, Axelrod and
Welch (2001). Struck and Willson (2002) have discussed the wake
dynamics that might be caused by a brown dwarf or planet orbiting within the
extended atmosphere of a mira variable, but they do not predict the type of
variation observed here.

Figure 7 shows the interesting case of a star (Blanco 26 =
BMB\footnote{Blanco, McCarthy and Blanco, 1985} 95) in which a
long-timescale variation of about 344d period seems to commence suddenly.
This phenomenon is not seen in other published light curves, in which the
long-period variability is always present. The first dip of Blanco 26, at
$\sim$ (JD - 2400000) + 49560, is merely a short event of a few days
duration. This brief and steep-sided drop in brightness seems to be
incompatible with a pulsational explanation for the light curve. It was
possibly caused by a short-lived outburst of obscuring material. The
following minimum, at 49900, again shows a sudden decline but is followed by
a slow recovery. The minima at 50250 and 50600 are almost symmetrical and
the last two observed show a shortening rise time. The periodic
occurrence of the dips suggests that the obscuring material may be in orbit,
shepherded by a compact star, or that its production may be stimulated at
the periastron of a binary system, possibly a precessing one. As is well
known, a loosely-bound cloud of particles with significant velocity
dispersion can be expected to spread with time along the orbit, and this
may explain the evolution of the phase-dependence of the obscuration with
time.

\begin{figure*}
\begin{minipage}{17.5cm}
\epsfxsize=17.5cm                                                          
\epsffile[28 495 539 786]{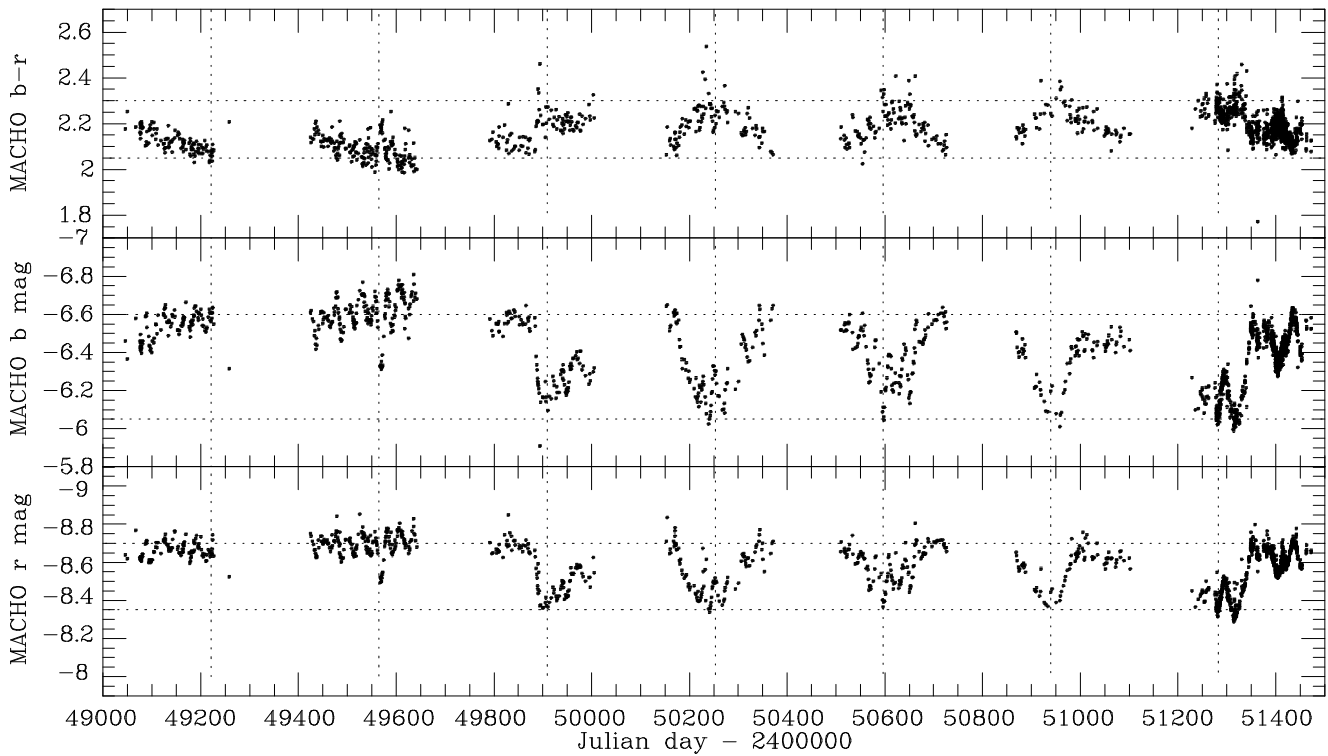}

\caption{MACHO light curves of Blanco 26 (=ISOGAL-PJ180323.0--300319), a
star in which a long-timescale variation ($\sim$344d) suddenly commences.
The instrumental magnitudes have been plotted; i.e., no zero-points have
been applied. The vertical dotted lines are placed at intervals of the
period. The short-lived event at JD $\sim$2449560 appears to be real, with
small errors in the MACHO data. The horizontal dotted lines are rough fits
by eye. These data are taken from a later release which covers seven
observing seasons.}

\end{minipage}
\end{figure*}                                                              

A short-period pulsation of about 27 days persists throughout the data and
can be seen in each season's data when analysed separately.

The $b-r$ colour of this star becomes redder when it is faint. This could be
the result of increased circumstellar obscuring material which, if it
resembles interstellar grains, is expected to have a greater effect at short
wavelengths than at long ones. Similar behaviour is seen in the light curves
of two other galactic double-period SRs, MACHO 113.18154.26 and
113.18285.65, illustrated in Alard et al (2001). The $r$ light curves of the
double-period SRVs in the LMC (Wood et al., 1999) show that the long periods
also persist throughout the interval over which the data are taken. The
variations are, in general, not sinusoidal and can sometimes be
sawtooth-shaped. The long-period minima can be short or long (up to $\sim
0.3$ times the period). A comparison with their $b$ light curves shows that
they also have colours which are redder when faint.

ISOGAL 15$\mu$m photometry of the galactic stars mentioned above was
obtained around JD\,2450876. Two, Blanco 26 and MACHO 113.18154.26, do not
show pronounced infrared excesses, whereas MACHO 113.18285.65 has [7] - [15]
= 1.3. In other words, there is no clear association of the double-period
phenomenon with high mass-loss. It is also possible that the obscuration
could be due to a TiO cloud in orbit. This possibility should be easy to
check by spectroscopy. The light minmima of RCB stars are believed to be
caused by the expulsion and rapid condensation of carbon vapour.

\section{Blanco stars and ISOGAL}
\subsection{ISOGAL objects in the Blanco field}

Glass et al (1999) found that nearly all the ISOGAL stars in the Blanco
field correspond to late-type M giants. The boundaries of the ISOGAL survey
fields are now more conservatively defined to avoid the unreliability
associated with their jagged edges.

The Blanco star positions were cross-correlated visually with the latest
version of the ISOGAL Catalogue, PSC1.0 (see Schuller et al, 2002), using
overlays. Somewhat more than half of the Blanco field overlaps the ISOGAL
NGC\,6522 field. About 70 Blanco objects lie outside. Thirty-seven Blanco
stars (including no.\ 54) were detected by ISOGAL, almost all of them
variables with sub-types of 5 or later (see Table 4 and Fig.\ 8). Five
Blanco stars detected by ISOGAL were non-variables.

As expected, the reddest stars in $K_S$ -- [15] and [7] -- [15] are associated
with the longest periods (log $P$ $>$ 1.8) and late spectral types (M5
and later). These objects have 15$\mu$m excesses almost certainly
attributable to dust.  However, there are also stars of late M types (M5, M6
and M6.5) which do not show 15$\mu$m excesses. The larger sample of stars
considered by Alard et al (2001) shows the trend more clearly. It should be
remembered that the ISOGAL and DENIS surveys were not simultaneous. However,
because the amplitudes of variation were small for the stars that we
are considering, the $K_S$ - [7] colours can be of value.

\begin{figure}                                                            
\epsfxsize=8.2cm                                                          
\epsffile[25 282 539 786]{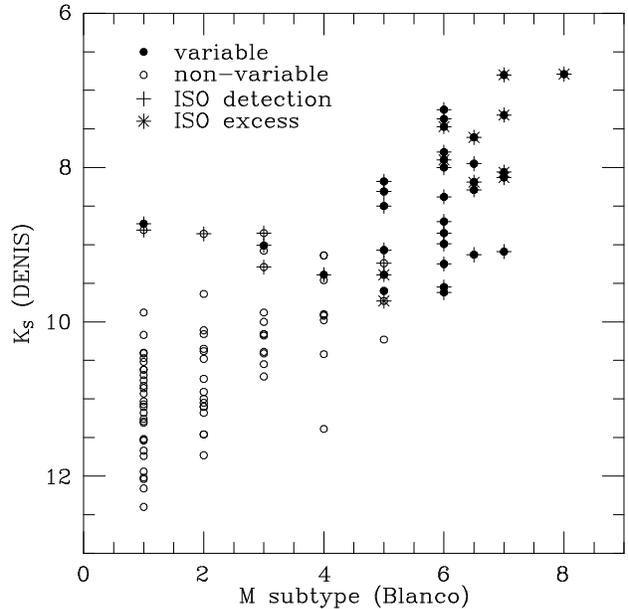}
\caption{DENIS $K_S$ vs spectral type for Blanco stars in the area common
with that covered by the ISOGAL PSC1.0 Catalogue. Stars detected by ISO are
distinguished as are those with [7] -- [15] $>$ 0.50 (ISO excess). The
excesses, implying high mass-loss rates, are confined to the most luminous
objects, having late spectral sub-types. The two non-variable M5 stars
appearing to possess excesses are probably marginal detections with high
expected errors at 15$\mu$m.}
\end{figure}                                                              

\begin{table}
\caption{ISOGAL Detection rate vs Blanco spectral sub-class (PSC1.0 
catalogue only)}
\begin{tabular}{lll}
Blanco class & detected & not detected\\
M1   & 1  & 33 \\ 
M2   & 1  & 16 \\
M3   & 3  & 10 \\
M4   & 4  & 4 \\
M5   & 5  & 2 \\
M6   & 12 & 0 \\
M6.5 & 5  & 0 \\
M7   & 4  & 0 \\
M8   & 1  & 0 \\

\end{tabular}

\noindent Note: The following Blanco objects are not counted in this table:
Blanco 2 (too close to Blanco 3 for ISOCAM to detect separately)
Blanco 54 (possible a composite object of M1 + something else)

\end{table}

\subsubsection{Notes on non-variables and individual stars}

The non-variables detected by ISOGAL were Blanco 22(4), 46 (4), 58(5), 86
(1) and 101 (4), where the spectral subtypes are given in brackets. The
detection of early sub-types is not common. These particular stars are among
the brightest members of their spectral sub-classes and close to or above
the threshold level for detection by ISOGAL at 7 $\mu$m. They may be
foreground objects or misclassified.

In addition to those discussed above, seven more Blanco stars not included
in the PSC1.0 Catalogue were detected by ISOGAL. They are situated at the
edge of the field where the data are less reliable. They are Blanco numbers
75, 84, 118, 126, 132, 151 and 154.

\subsubsection{Non-Blanco ISOGAL stars}

Forty-eight ISOGAL sources were within the annulus specified by Blanco and
nine of these were not listed by Blanco.  Of these nine,
ISOGAL-PJ180340.6--300002 and ISOGAL-PJ180344.2--295845 are not detected at
7$\mu$m or at the DENIS wavelengths and are faint on the Blanco chart. They
are also near the limit of detection at 15$\mu$m and may not be real.

\begin{table}
\caption{Non-Blanco ISOGAL detections}
\begin{tabular}{llllll}
ISOGAL-PJ       &  $I$   & $J$   & $K_S$ & [7]  & [15] \\
180320.7-300138 &  12.44 & 10.95 & 10.04 & 9.22 &      \\
180323.6-300159 &  11.71 & 10.18 & 9.02  & 8.57 &      \\
180326.8-300131 &  11.46 & 10.30 & 9.02  & 9.29 &      \\
180327.4-300240 &  12.08 & 10.52 & 9.56  & 9.34 &      \\
180328.7-300356 &  12.0  & 10.68 & 9.55  & 9.16 &      \\
180328.9-300335 &  10.71 & 9.42  & 8.51  & 8.41 &      \\
180329.6-300109 &  12.31 & 9.51  & 7.96  & 7.42 & 6.84 \\
\end{tabular}                                                
\end{table}

The remaining seven (see Table 5) are bright on the Blanco chart.
ISOGAL-PJ180329.6--300109 was detected at both ISOGAL wavelengths. 
Its MACHO light curve shows that it is a SRV with period 83d and amplitude
0.4 mag in $r$. Its DENIS colours suggest a spectral type around Blanco M5.
ISOGAL-PJ180323.6-300159 is also a SRV, with P=34d and amplitude
$\sim$0.05. In this case, the DENIS colours suggests a Blanco type about
M2--3. Two other stars have infrared colours corresponding to early Blanco
M-type. They are ISOGAL-PJ180326.8-300131 (noisy MACHO trace) and
ISOGAL-PJ180328.7-300356, which do not show significant variability. 

The other three objects were detected at 7 but not at 15$\mu$m.
ISOGAL-PJ180328.8-300335, is a photometric standard star mentioned by
Arp (1965), with $V$=12.74, $B-V$=0.53.

A clear implication of the discovery of four new probable M stars in this
field is that the Blanco (1986) survey is not quite complete. These stars
are among the brightest at $I$ in the field and may have been saturated on
Blanco's grism plate. In some cases there may have been confusion due to
crowding of images.

\subsubsection{Variable Blanco star not detected in ISOGAL}

Only one Blanco star that is variable and that falls within the ISOGAL area
was {\it not} detected, viz.\ 83 (5). Blanco 83 has an unusually regular
light curve with a 135.5d period, and may not be a typical SR variable. Its
DENIS colours $(I-J, J-K_S)_0$ = (1.62, 1.15) are however compatible with its
spectral type.

\section{Conclusions}

We have examined a `complete' sample of 174 Blanco M giants in the Inner
Bulge of the Galaxy for variability and infrared characteristics.

About 46 of the Blanco (1986) stars are semi-regular variables with
amplitudes exceeding about 0.03 mag. Stars with spectral types of M6 or
later are variable, together with many of type M5, especially the most
luminous. More than one period is fairly common.

There is a continuous increase of average luminosity and of $I-J$ and $J-K_S$
colours with M sub-class. A $K_S$, log $P$ trend is shown by SRVs if in each
case the most important shorter period is used when a long period is present.

Thirty-seven Blanco stars out of a possible $\sim$101 were detected by ISO
during the ISOCAM programme. It is clear that there is a high correlation
between variability and detection but not whether this is due to luminosity
or to temperature effects (late spectral type). NGC\,6522 therefore contains
very few semi-regular variable M-type giants that escaped detection in the
work of Alard et al (2001). 

Four additional probable M stars detected by ISO, two of them
semi-regular variables with late M-subtype colours, have been found in
the Blanco (1986) field, which shows that his survey is not as complete as
formerly believed.

Among those stars with both short and long periods, the long periods appear
to fall close to the `D' line in the $K_S$, log $P$ diagram,  found by Wood
(2000) for similar objects in the LMC. The unusual light curve for Blanco 26
suggests that pulsation cannot be responsible for the long period
variability. 

\section{Acknowledgments}

The STAR programs written by Dr L.A. Balona (SAAO) were used in the Fourier
analysis of the light curves.

We thank Dr C. Koen (SAAO) and Dr A. Omont (IAP) for reading drafts of this
paper and suggesting some improvements.

ISG thanks Prof S. Deguchi and the National Astronomical Observatory of
Japan as well as Prof Y. Nakada and the Institute of Astronomy, University
of Tokyo, for their hospitality during part of this work.

MS is supported by the Fonds zur F\"{o}rderung der Wissentschaftlichen
Forschung (FWF), Austria, under project J1971-PHY. 

This paper utilizes public domain data originally obtained by the MACHO
Project, whose work was performed under the joint auspices of the U.S.
Department of Energy, National Nuclear Security Administration by the
University of California, Lawrence Livermore National Laboratory under
contract No. W-7405-Eng-48, the National Science Foundation through the
Center for Particle Astrophysics of the University of California under
cooperative agreement AST-8809616, and the Mount Stromlo and Siding Spring
Observatory, part of the Australian National University.

We acknowledge use of the Digitized Sky Survey produced at the Space
Telescope Science Institute under US Government grant NAG W-2166, based on
material taken at the UK Schmidt Telescope, operated by the Royal
Observatory Edinburgh with funding from the UK Science and Engineering
Research Council and later by the Anglo-Australian Observatory.

\end{document}